\documentclass[prd,showpacs,reprint,longbibliography]{revtex4-1}

\usepackage{epsfig,amsmath,amssymb}
\usepackage[outdir=./]{epstopdf}

\begin{document}
	
\title{Spin Gravitational Resonance and Graviton Detection} 

\author{James Q. Quach} 
\email{quach.james@gmail.com}
\affiliation{Institute for Solid State Physics, University of Tokyo, Kashiwa, Chiba 277-8581, Japan}

\begin{abstract}
We develop a gravitational analogue of spin magnetic resonance, called spin gravitational resonance, whereby a gravitational wave interacts with a magnetic field to produce a spin transition. In particular, an external magnetic field separates the energy spin states of a spin-1/2 particle, and the presence of the gravitational wave produces a perturbation in the components of the magnetic field orthogonal to the gravitational wave propagation.  In this framework we test Dyson's conjecture that individual gravitons cannot be detected. Although we find no fundamental laws preventing single gravitons being detected with spin gravitational resonance, we show that it cannot be used in practice, in support of Dyson's conjecture.
\end{abstract}

̀
%04.60.Bc	Phenomenology of quantum gravity
%04.30.-w	Gravitational waves
%03.65.Pm	Relativistic wave equations
%04.62.+v	Quantum fields in curved spacetime
%03.65.Sq	Semiclassical theories and applications

\maketitle

\section{Introduction}

The recent detection of gravitational waves (GWs) of coalescing binary black holes by LIGO \cite{abbott16} was the first direct measurement. This monumental event puts the existence of GWs beyond doubt, ushering in a new era of GW experiments. This leads to the question of whether gravitons, the theorised quantum carrier of the gravitational force, can be detected, even in principle. Dyson has conjectured that no conceivable experiment in our universe can detect a single graviton~\cite{dyson04}. He distinguishes the proving of the existence of physical laws which prevent the detection of single gravitons from the argument that the detection of single gravitons is undetectable in practice. In the context of the latter, Dyson argued that detectors based on LIGO, interaction with individual atoms, and coherent transitions between graviton and photon states, cannot be used to detect gravitons in practice \cite{dyson13}. In two papers Boughn and Rothman~\cite{boughn06,rothman06} investigated Dyson's conjecture and came to the conclusion that although there was no fundamental reason forbidding the detection of a single graviton, in practice it would be impossible. In their work they considered the ionization and state transition of hydrogen atoms by high-energy gravitons, neglecting spin.

Recently the correct non-relativistic limit of the  Dirac Hamiltonian in a singularly polarised GW background was derived~\cite{quach15}. Here we generalise this Hamiltonian to arbitrary polarisation and determine whether the inclusion of the spin degree of freedom can change the conclusion that in practice single gravitons are not detectable. Specifically, we look at the absorption of gravitons in spin states split by an external magnetic field through the Zeeman effect. The process is analogous to spin magnetic resonance, and so we call it spin gravitational resonance (SGR).

In Sec.~\ref{sec:dirac} we write down the non-relativistic limit of the  Dirac Hamiltonian in a generally polarised GW background. In Sec.~\ref{sec:sgr} we develop the theory of SGR. In Sec.~\ref{sec:sources} we consider whether SGR could ever be used to detect GWs and single gravitons.

\begin{figure}
	\centering
	\includegraphics[width=.8\columnwidth]{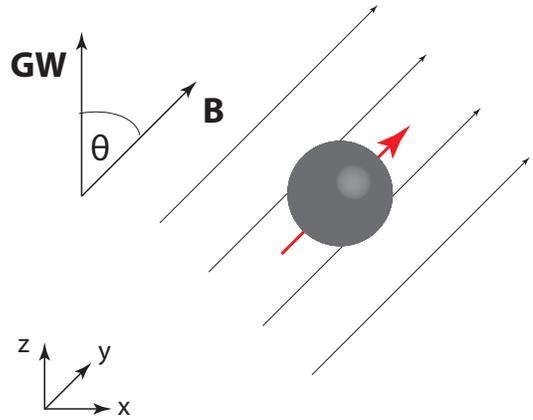}
	\caption{(color online) A constant external magnetic field $\mathbf{B}$ (depicted here lying in the $x\!-\!z$ plane) lifts the degeneracy of spin states. In the \textit{ground state}, the particle's spin (red arrow) is aligned with the magnetic field. When the propagation direction of the plane GW is not (anti-)parallel to the magnetic field, the interaction of the magnetic field with the GW in the plane perpendicular to its propagation produces the perturbation that results in the transition between spin states. }
	\label{fig:gw}
\end{figure}

\section{Non-relativistic limit of the Dirac Hamiltonian in a generally polarized gravitational-wave background}
\label{sec:dirac}

The Dirac equation in curved spacetime is,
\begin{equation}
	i\hbar\gamma_ae_\mu^a(\partial_\mu-\Gamma_\mu-\frac{ie}{\hbar}A_\mu)\psi=mc\psi~.
\label{dirac}
\end{equation}
where  $\gamma_a$ are gamma matrices defining the Clifford algebra $\{\gamma_a,\gamma_b\}=-2\eta_{ab}$, with spacetime metric signature ($-,+,+,+$). $e_\mu^a$ are the tetrads that relate at every point the metric $g_{\mu\nu}$ to a tangent Minkowski space via $g_{\mu\nu}=e_\mu^a e_\nu^b\eta_{ab}$. The spinorial affine connection $\Gamma_\mu=\frac{i}{4} e_\nu^a (\partial_\mu e^{\nu b}+\Gamma_{\mu \sigma}^\nu e^{\sigma b}) \sigma_{ab}$, where $\Gamma_{\mu\sigma}^\nu$ is the affine connection and $\sigma_{ab}\equiv\frac{i}{2}[\gamma_a,\gamma_b]$ are the generators of the Lorentz group.  $A_\mu$ is the electromagnetic four-vector potential. The Einstein summation convention where repeated indices ($\mu,\nu,\sigma,a,b=\{0,1,2,3\}$) are summed, has been used.

The metric for a generally polarised linear plane GW is
\begin{equation}
	ds^2=-c^2dt^2+dz^2+(1-2v)dx^2+(1+2v)dy^2-2udxdy~,
\label{gw_metric}
\end{equation}
where $u=u(t-z)$ and $v=v(t-z)$ are functions which describe a wave propagating in the $z$-direction. We will consider the case of a circularly polarised GW travelling along the $z$-direction, i.e. $v=f=f_0e^{i(kz-\omega t)}$ and $u=if$. Under this metric Eq.~(\ref{dirac}) can be written in the familiar Schr\"{o}dinger picture $i\hbar\partial_t \psi=H\psi$, where ($\boldsymbol{\alpha}\equiv\gamma^0\boldsymbol{\gamma},\beta\equiv\gamma^0,\boldsymbol{p}\equiv-i\hbar\nabla$, and indices $i,..,n=\{1,2,3\}$~\cite{goncalves07}
\begin{equation}
	H=\beta mc^2+c\alpha^j(\delta_j^i+T_j^i)(p_i-eA_i)~,
\label{eq:H}
\end{equation}
with
\begin{equation}
	T=
		\begin{pmatrix}
			v& &-u &0\\
			-u& &-v &0\\
			0& &0 &0\\
		\end{pmatrix}~.
\label{eq:T}
\end{equation}

A means by which to write down the non-relativistic limit of the Dirac Hamiltonian with relativistic correction terms is provided by the Foldy-Wouthuysen (FW) transformation~\cite{foldy78}. The FW transformation is a unitary transformation which separates the upper and lower spinor components. In the FW representation, the Hamiltonian and all operators are block-diagonal (diagonal in two spinors). There are two variants of the FW transformation known as the \textit{standard} FW (SFW)~\cite{foldy78} and \textit{exact} FW (EFW)~\cite{eriksen58,nikitin98,obukhov01,jentschura14} transformations. 

\subsection{Exact Foldy-Wouthuysen Transformation}

Central to the EFW transformation is the property that when $H$ anti-commutes with $J\equiv i\gamma^5 \beta$, $\{H,J\}=0$, under the unitary transformation $U=U_2 U_1$, where ($\Lambda\equiv H/\sqrt{H^2}$) 
\begin{equation}
U_1=\frac{1}{\sqrt{2}} (1+J\Lambda),\quad\quad U_2=\frac{1}{\sqrt{2}} (1+\beta J)~,
\end{equation}
the transformed Hamiltonian is even (even terms do not mix the upper and lower spinor components, odd terms do), 
\begin{equation}
\begin{split}
UHU^+=&\frac12\beta(\sqrt{H^2}+\beta\sqrt{H^2}\beta)+\frac12(\sqrt{H^2}-\beta\sqrt{H^2}\beta)J\\
=&\{\sqrt{H^2}\}_\text{even}\beta+\{\sqrt{H^2}\}_\text{odd}J~.
\end{split}
\label{UHU}
\end{equation}
Note that as $\beta$ is an even operator and $J$ is an odd operator, Eq.~(\ref{UHU}) is an even expression which does not mix the positive and negative energy states.

Our Hamiltonian satisfies the EFW anti-commutation property. Neglecting small $T^{ij}T^{lm}$ order terms and using the identity $\alpha^i\alpha^{j}=i\epsilon^{ijk}\sigma_k \textbf{I}_{2} + \delta^{ij}\textbf{I}_{4}$, the perturbative expansion of $\sqrt{H^2}$ yields to $O[1/m]$ accuracy ($\sigma_i$ are Pauli matrices),
\begin{equation}
\begin{split}
H_\text{EFW}=&\frac{1}{2m}(\delta^{ij}+2T^{ij})[(p_i-eA_i)(p_j-eA_j)]\\
&+\frac{e\hbar}{4m}(\delta^{ij}+2T^{ij})\epsilon_{jkl}\sigma^l[\partial^k(A_i)-\partial_i(A^k)]\\
&+\frac{\hbar}{2m}\partial^i(T^{jl})\epsilon_{ijk}\sigma^k(p_l-eA_l)+mc^2~.
\end{split}
\label{H_FW}
\end{equation}

Note that $\sqrt{H^2}=\{\sqrt{H^2}\}_\text{even}=H_\text{FW} \textbf{I}_{2}$ contains only even terms, and therefore $\{\sqrt{H^2}\}_\text{odd}=0$ in Eq.~(\ref{UHU}). Eq.~(\ref{H_FW}) generalises the non-relativistic limit of the Dirac Hamiltonian of a singular polarised GW background derived in Ref.~\cite{quach15}, to arbitrary polarisation. As a check of the correctness of Eq.~(\ref{H_FW}), we will also show that application of the SFW will also result in Eq.~(\ref{H_FW}). 

The first and second terms of $H_\text{FW}$ involves the kinetic and magnetic dipole energies and their corrections due to the GW. Ref.~\cite{boughn06,rothman06,dyson13} only used the first term to calculate the probability of graviton detection, neglecting the effects of spin. Here we will consider the second term to calculate the effects of spin on graviton absorption. The third term can be thought of as being the GW analogue of the Schwarzschild gravitational spin-orbit energy~\cite{oliveira62}. At atomic dimensions, the gravitational wavelength is long, and this term can be considered negligible. The last term is the rest mass energy.

We note that by beginning with the minimally coupled Dirac equation [Eq.~({\ref{dirac}})] where the gauge field is coupled through the covariant derivative, we do not account for the anomalous magnetic moment (AMM). A non-minimal coupling is required to account for the AMM~\cite{morishima04,obukhov14}. In the non-relativistic limit the effect of the AMM amounts to multiplying the magnetic moment by $g/2\approx1.001$. This is a small effect which will bear no qualitative influence on our conclusions.

\subsection{Standard Foldy-Wouthuysen Transformation}

In this section for convenience we will work in the natural units where $\hbar=c=e=1$. We will put $\hbar,c,e$ back into the final equation. 

The odd and even components of $H$ are respectively given by,
\begin{equation}
\mathcal{O}=\frac12(H-\beta H\beta),\quad \mathcal{E}=\frac12(H+\beta H\beta)~.
\label{odd_even}
\end{equation}

In the SFW method we use the unitary transformation $U=e^{iS}$, where $S=-\frac{i\beta}{2m}\mathcal{O}$. Applying the transformation through $\psi'=e^{iS}\psi$, the Schr\"{o}dinger equation becomes,
\begin{equation}
	i\partial_t \psi' = [e^{iS}(H-i\partial_t)e^{-iS}]\psi'=H'\psi'
\end{equation}
The transformed Hamiltonian $H'=e^{iS}(H-i\partial_t)e^{-iS}$ is then expanded in a series of multiple commutators using the Baker-Hausdorff lemma,
\begin{equation}
	e^{i\lambda S}Me^{-i\lambda S}=M+i\lambda [S,M]+\frac{(i\lambda)^2}{2!}[S,[S,M]]+\cdots~,
\end{equation}
This gives,
\begin{equation}
H'\approx H+i[S,H]-\frac{1}{2}[S,[S,H]]-\dot{S}-\frac{i}{2}[S,\dot{S}]~.
\label{H_prime}
\end{equation}
$i[S,H]\approx-\mathcal{O}$ generates a term that eliminates the odd operator $\mathcal{O}$, however many more terms are generated by the higher-order terms which potentially could be odd operators. To eliminate these odd operators, the FW transformation is repeated on subsequent Hamiltonians (i.e. $H',H'', H'''$, and so on) until all odd operators are eliminated to the required order of accuracy.  In the non-relativistic limit the rest mass energy dominations, and the SFW provides an expansion in increasing accuracy in powers of $1/m$. 

Repetition of the SFW transformation three times gives to O[$1/m$] accuracy~\cite{bjorken64}
\begin{equation}
	H'''=\beta\Bigl(m+\frac{\mathcal{O}^2}{2m}\Bigr)~,
\label{eq:H'''}
\end{equation}
where $\mathcal{O}=c\alpha^j(\delta_j^i+T_j^i)\pi_i$ with $\pi_i\equiv p_i-eA_i$. Thus one only needs to compute $\mathcal{O}^2$, with the result that ($\Sigma_{k}\equiv \sigma_k \textbf{I}_2$)
\begin{align}
	\mathcal{O}^2&=\pi_i\pi^i-i\epsilon^{ijk}\Sigma_k p_i(A_j)+(\delta^{ij}+i\epsilon^{ijk}\Sigma_k)p_i(T_j^l)\pi_l\nonumber\\
		&+T^{ij}(\pi_j\pi_i+\pi_i\pi_j)+i\epsilon^{ijk}\Sigma_kT_j^l[p_l(A_i)-p_i(A_l)]~.
\label{eq:O^2}
\end{align}
In Eq.~({\ref{eq:O^2}}) we have as usual omitted the small $T^{ij}T^{kl}$ term. Substituting Eq.~({\ref{eq:O^2}}) into Eq.~({\ref{eq:H'''}}) and explicitly reinstating $\hbar,c,e$, one gets
\begin{equation}
\begin{split}
	H'''=&\frac{\beta}{4m}(\delta^{ij}+2T^{ij})[(\pi_i\pi_j+\pi_j\pi_i]\\
		&+\frac{\beta e\hbar}{4m}(\delta^{ij}+2T^{ij})\epsilon_{jkl}\Sigma^l[\partial^k(A_i)-\partial_i(A^k)]\\
		&+\frac{\beta\hbar}{2m}(\epsilon_{ijk}\Sigma^k-i\delta_{ij})\partial^i(T^{jl})\pi_l+\beta mc^2~.
\end{split}
\label{eq:H_SFW}
\end{equation}
Using the fact that $\delta_{ij}\partial^i(T^{jl})=0$ and $T$ is symmetrical, we retrieve Eq.~(\ref{H_FW}) from Eq.~(\ref{eq:H_SFW}). In other words $H_\text{EFW}=H_\text{SFW}$ to $O[1/m]$ accuracy, where $H'''=\beta H_\text{SFW}$. The EFW and SFW transformations are not equivalent unitary transformations, and in general can give rise to different Hamiltonians in the non-relativistic limits. The exact relationship between the the EFW and SFW is given by Ref.~\cite{silenko05,neznamov09}. In the current case however, that both the EFW and SFW transformation yields the same Hamiltonian, is good verification that Eq.~(\ref{H_FW}) is the correct non-relativistic limit of the Dirac Hamiltonian in the presence of an EM gauge and GW field to O[$1/m$] accuracy. Note that the EFW and SFW transformation may yield differing higher order correction terms beyond O[$1/m$] accuracy.

\section{Spin gravitational resonance}
\label{sec:sgr}

Spin magnetic resonance, of which nuclear magnetic resonance and electron spin resonance (ESR) are examples of, has found widespread use, ranging from magnetic resonance imaging to ESR spectroscopy. The underlying physics is that a constant external magnetic field lifts the degeneracy of spin states. The transition between the states occur on the absorption of a photon from an external oscillating electromagnetic field. In an analogous manner we propose the theoretical possibility of GWs interacting with a constant external magnetic field to excite this transition, i.e. SGR.

We consider the case of a spin-$1/2$ particle in a constant magnetic field $\mathbf{B}$, perturbed by a circularly polarised GW. The direction of the magnetic field and GW propagation is separated by small angle $\theta$ as shown in Fig.~(\ref{fig:gw}). As noted previously the gravitational wavelength is much larger than atomic dimensions so that $\partial(T)$ is small. Neglecting this small term, the spin component of the Hamiltonian is the second term of Eq.~(\ref{H_FW}) ($\mu_B\equiv\frac{e\hbar}{2m}$),
\begin{equation}
	H_\text{S}=\frac{\mu_B}{2}(\delta^{ij}+2f\mathcal{T}^{ij})\epsilon_{jkl}\sigma^l[\partial^k(A_i)-\partial_i(A^k)]~,
\label{H_I}
\end{equation}
where
\begin{equation}
	\mathcal{T}=
		\begin{pmatrix}
		1& &-i &0\\
		-i& &-1 &0\\
		0& &0 &0\\
		\end{pmatrix}~.
\end{equation}

We rewrite Eq.~(\ref{H_I}) with $\epsilon^{mki}B_m\equiv\partial^k(A_i)-\partial_i(A^k)$ and use the identity $\epsilon_{ijk}\epsilon^{imn}=\delta_j^m\delta_k^n-\delta_j^n\delta_k^m$; after taking the real components one gets
\begin{equation}
\begin{split}
	H_\text{S}=&-\mu_B\boldsymbol{\sigma}\cdot\mathbf{B}+\mu_Bf_0[(\sigma^xB_x-\sigma^yB_y)\cos(kz-\omega t)\\
&+(\sigma^xB_y+\sigma^yB_x)\sin(kz-\omega t)]~.
\end{split}
\label{eq:H_S}
\end{equation}

From Eq.~(\ref{eq:H_S}) one identifies $\mathbf{B}$ as the magnetic field in the absence of a GW, and the negative sign in front of the first term reminds us that the \textit{ground state} of the system occurs when the particle's spin is aligned with the magnetic field. 

Without loss of generality we take the magnetic field  to be rotated about the $y$-axis as shown in Fig.~(\ref{fig:gw}), $\textbf{B}=B(\sin\theta,0,\cos\theta)$. We would like now to write the spin operator in a co-ordinate basis aligned with the direction of the magnetic field. Using the Baker-Hausdorff lemma, the spin operators transform as (prime indicates operators in the co-ordinate basis aligned with the magnetic field)
\begin{equation}
\begin{split}
	\sigma_x&\rightarrow e^{i\theta \sigma_y'/2}\sigma_x'e^{-i\theta \sigma_y'/2}\nonumber\\
	&=\sigma_x'+\frac{i\theta}{2}[\sigma_y',\sigma_x']+\frac{1}{2!}\Bigl(\frac{i\theta}{2}\Bigr)^2[\sigma_y',[\sigma_y',\sigma_x']]+\cdots\nonumber\\
	&=\sigma_x'(1-\frac{\theta^2}{2!}+\cdots)+\sigma_z'(\theta-\frac{\theta^3}{3!}+\cdots)\nonumber\\
	&=\sigma_x'\cos\theta+\sigma_z'\sin\theta~,	
\end{split}
\label{eq:S_x}
\end{equation}

Together with similar transformations for $\sigma_y,\sigma_z$, the transformation of the spin operators are summarised in the following,
\begin{equation}
	\boldsymbol{\sigma}\rightarrow
		\begin{pmatrix}
		\cos\theta& &0 &\sin\theta\\
		0& &1 &0\\
		-\sin\theta& &0 &\cos\theta\\
		\end{pmatrix}~\boldsymbol{\sigma}'~.
\end{equation}

In terms of $\boldsymbol{\sigma}'$ the Hamiltonian is,
\begin{equation}
\begin{split}
	H_\text{S}=&-\mu_BB\sigma_z'+\mu_Bf_0B[\sin\theta\cos\theta\cos(kz-\omega t)\sigma_x'\\
	&+\sin\theta\sin(kz-\omega t)\sigma_y'+\sin^2\theta\cos(kz-\omega t)\sigma_z']~.
\end{split}
\label{eq:H_I_basis}
\end{equation}

In the small $\theta$ limit, this becomes,
\begin{equation}
	H_\text{S}\approx-\mu_B B\sigma_z'+\frac12\mu_Bf_0B\theta[e^{i(kz-\omega t)}\sigma_-+e^{-i(kz-\omega t)}\sigma_+]
\label{eq:H_I_limit}
\end{equation}
where $\sigma_\pm\equiv\sigma_x'\pm i\sigma_y'$ are the spin lowering and raising operators. The small angle limit allows a closed-form solution, without qualitatively changing the outcome.

One can identify Eq.~(\ref{eq:H_I_limit}) as a Rabi model for a driven coupled two level quantum system,  with the probability of finding the system in the \textit{excited state}  given by the Rabi formula~\cite{jaynes63},
\begin{equation}
	P_+=\Bigl(\frac{\gamma}{\Omega}\sin\Omega t\Bigr)^2
\label{eq:P+}
\end{equation}
where $\gamma\equiv\mu_Bf_0B\theta/2\hbar$, $\epsilon=\mu_BB/\hbar$ is the difference in energy between the two states, and Rabi frequency $\Omega=\sqrt{\gamma^2+(\omega-\epsilon)^2/4}$. We have set the system to be in the ground state at $t=0$. The probability of the system being in the ground state is given by $P_-=1-P_+$~. Eq.~(\ref{eq:P+}) provides an analytical form for the probability of a GW exciting a two-level spin-$1/2$ system in a constant magnetic field for small $\theta$, and is the main outcome of this section.

The interaction of the magnetic field with the oscillation of the GW in the plane perpendicular to its propagation produces the perturbation that results in the transition of spin states. If the GW propagation was parallel to the magnetic field this would not perturb the magnetic field and therefore there would be no state transition. This can be seen by setting $\theta=0$ in Eq.~(\ref{eq:H_I_basis}) which gives $H_S({\theta=0})=-\mu_BB\sigma_z'$. Perpendicular perturbation of the magnetic field arises when the magnetic field and GW propagation direction are not aligned (or anti-aligned), and therefore the possibility of a state transition. 

\section{Gravitational Wave Sources}
\label{sec:sources}

The tunable small energy gap between the spin states allows for the absorption of low frequency gravitational sources. The amplitude of the probability of absorption is not dependent on the source frequency, but on the difference between the gravitational source and resonant frequency of the spin system - the maximum amplitude of 1 occurs when the source frequency equals the resonant frequency. Although the amplitude of the probability of absorption is independent of the source frequency, the time it takes to reach the maximum probability is dependent on the source frequency. In this section we look at the case of resonant low frequency astronomical and local gravitational sources. We begin by considering whether SGR can be used to detect GWs first before considering whether it can be used to detect single gravitons.

Astronomical binary systems offer a continuous source of GWs. Prior to the recent direct detection of a binary black hole merger, the PSR B1913+16 binary pulsar system provided the strongest indirect evidence of GWs~\cite{hulse75,taylor82}.  For two equal masses $m$ in circular orbit about each other with angular velocity $\alpha$, and separated by distance $l$, they will generate circularly polarized GWs with frequency $\omega=2\alpha$ and amplitude~\cite{schutz09}
\begin{equation}
	f_0=2G ml^2\omega^2/c^4r~.
\label{eq:f0_schutz}
\end{equation}

The PSR B1913+16 binary pulsar system is estimated to consists of two neutron stars of equal mass, $m=1.4~\text{M}_\odot$, with an orbital period of 7 h 45 min 7 s, at a distance of 8 kpc away. At this distance this binary system produces GWs with an amplitude of $f_0=10^{-23}$ on Earth with frequency $\omega=2.4\times10^{-4}$ s$^{-1}$. 

From Sec.~\ref{sec:sgr} we see that for a resonant transition ($\omega=\epsilon$), the magnetic field strength needs to be $B=\omega\hbar/\mu_B\approx10^{-15}$ T. Given that the Earth's magnetic field is on the order of $10^{-5}$ T, the experiment would need to be conducted inside a magnetic shield or in space. If the magnetic field in our graviton detection experiment was set up so that $\theta=0.1$,  the Rabi frequency would be on the order of $10^{-28}$ s$^{-1}$. If we took a typical detector density of $\rho=10^{25}$ atoms per m$^3$, and run the experiment for 1 year (yr), the number of gravitons detected would be $N=\rho P_+\approx10^{-17}$ m$^{-3}$, using Eq.~(\ref{eq:P+}). In other words, in any realistic experiment we would expect to detect no gravitons (or GWs).

Alternatively we could generate GWs in the laboratory with two locally orbiting masses. For two $m=1$ tonne masses separated by distance $l=100$ m orbiting at $\alpha=10^4$ s$^{-1}$, the generated GW would have $\omega=2\times10^4$ s$^{-1}$ with $f_0\approx 10^{-29}$, 1 m away from the source. The required magnetic field strength is $B=10^{-7}$ T, and therefore the experiment should also be shielded from the Earth's magnetic field. The number of gravitons detectable would be $N\approx10^{-11}$ m$^{-3}$ per year. Again not one single graviton would likely be detected. Note that the generator of the GW would be a formidable technological challenge, as it involves two 1 tonne masses travelling at $10^6$ m/s. The energy required to run it for one year would be on the order of 10$^{16}$ J, the amount required to run a small country (such as Mongolia) for one year.

The Rabi oscillations are particularly slow at $\Omega=10^{-28}$~s$^{-1}$ for the case of the binary pulsar system, and $\Omega=10^{-25}$~s$^{-1}$ for the terrestrial binary masses. A complete Rabi cycle would take longer than the age of the Universe. High frequency gravitational sources could reduce these times. However a relative abundance of high frequency gravitons is difficult to come by terrestrially - Refs.~\cite{rothman06,dyson13} discuss some potential sources. Furthermore,  a crucial problem with detecting high energy gravitons is discriminating them from the abundance of high energy solar neutrinos. The cross-section of high energy neutrino interaction with matter is at least 20 orders of magnitude larger than gravitons of the same energy on Earth~\cite{rothman06,dyson13}. The amount of ordinary material required to shield from the solar neutrinos would amount to light years, and would in fact collapse into a black hole under its own weight~\cite{rothman06}. 

In our considerations so far, we have limited ourselves to near-Earth detectors. Let us remove this restriction and propose we have a detector 1 AU away from the PSR B1913+16 binary pulsar system. At this distant the GW amplitude is $f_0=10^{-14}$, and the detection of the GW may be possible with $N\approx10$ m$^{-3}$ after one year.

The analogue of the Poynting flux or intensity for the gravitational plane wave is $I=c^3\omega^2f_0^2/8\pi G$~\cite{misner73,boughn06,rothman06}. For a GW amplitude of $f_0=10^{-14}$ at frequency $\omega=2.4\times10^{-4}$ s$^{-1}$, the density of gravitons would be $I/c\hbar\omega\approx 10^{28}$ m$^{-3}$. At this density, although we have detected a classical GW, we cannot claim to have detected individual gravitons.

In the classical picture, if the intensity of a classical wave is reduced such that its energy over a local volume is less than the energy that separates two states, the transition from the lower energy state to the higher one cannot occur. In the quantum picture, the energy of a particle is proportional to its frequency, independent of intensity. Therefore even if the intensity is reduced such that classically a state transition should not occur, the quantum picture can still allow a transition. This is the basis of the photoelectric effect. If we reduce the intensity of the GW source so its energy over the volume of the detector is less than the transition energy, a classical picture would suggest that there is not enough energy to affect a transition. If a transition were detected it would be difficult to reconcile this with a classical picture of the GW, suggesting a quantum mechanical nature. For the GW energy in a detector of volume $V$ to be less than the transition energy at resonance ($IV/c<\hbar\omega$), the amplitude of the GW needs to satisfy 
\begin{equation}
	 f_0<\sqrt\frac{{8\pi G\hbar}}{\omega Vc^2}~.
\label{eq:f0}
\end{equation}
Eq.~(\ref{eq:f0}) is a general condition for the detection of a single graviton, independent of the source of the GW and the details of the detector.

For our $V=1~\text{m}^3$ detector,  Eq.~(\ref{eq:f0}) requires that $f_0<10^{-28}$. At this amplitude, $N<10^{-26}$ m$^{-3}$ after one year. Therefore SGR cannot be used to detect single low frequency gravitons. In fact at resonance,  (from $N=\rho P_+$) $f_0=2\arcsin(\sqrt{N/\rho})/\theta\omega t$, and Eq.~(\ref{eq:f0}) implies that,
\begin{equation}
	\omega>\frac{c^2V\arcsin^2\sqrt{N/\rho}}{2\pi G\hbar\theta^2 t^2}~.
\label{eq:omega}
\end{equation}

If we want to detect at least a single graviton ($N=1~\text{m}^{-3}$) after one year with our SGR detector, then $\omega>10^{23}~\text{s}^{-1}$. In other words we would need a high frequency GW source. However as discussed in Ref.~\cite{boughn06,rothman06,dyson13}, known high frequency gravitational sources are not practical for single graviton detection.

\section{Conclusion}
Dyson conjectured that single gravitons cannot be detected in practice. Subsequent investigations into the ionisation and state transition of the hydrogen atom without spin supported the conjecture. We investigated the interaction of GWs with the spin degree of freedom of spin-$1/2$ particles in an external magnetic field and developed the theoretical framework for SGR. In doing so, although we find no fundamental laws preventing individual graviton detection, we showed that SGR cannot be used to detect single gravitons in practice, and therefore Dyson's conjecture is not violated, even when one accounts for spin.

\section*{Acknowledgements}

The author would like to thank M. Lajk\'{o}, C.-H. Su, I. Yaakov, and S. Quach for discussions and checking the manuscript. This work was financially supported by the Japan Society for the Promotion of Science. The author is an International Research Fellow of the Japan Society for the Promotion of Science.

\bibliography{sgr}

\end{document}